\documentclass[fleqn,usenatbib]{mnras}

\usepackage{newtxtext,newtxmath}

\usepackage[T1]{fontenc}
\usepackage{ae,aecompl}

\usepackage{nicefrac}

\usepackage{graphicx}	
\usepackage{amsmath}	
\usepackage{amssymb}	

\title{The TESS light curve of AI Phoenicis
}

\author[P.~F.~L.~Maxted et al.]{
P.~F.~L.~Maxted,$^{1}$\thanks{E:mail: p.maxted@keele.ac.uk, gaulme@mps.mpg.de,
darek@astro-udec.cl, xysiek@ncac.torun.pl, colecampbell.johnston@kuleuven.be,
jorosz@sdsu.edu, aprsa@villanova.edu, astro.js@keele.ac.uk, gtorres@cfa.harvard.edu, G.R.Davies@bham.ac.uk}
Patrick~Gaulme,$^{2}$\footnotemark[1]
D.~Graczyk,$^{3}$\footnotemark[1]
K.~G.~He\l miniak,$^{3}$\footnotemark[1]
\newauthor
C. Johnston,$^{4}$\footnotemark[1]
Jerome~A.~Orosz,$^{5}$\footnotemark[1]
Andrej Pr\v{s}a,$^{6}$\footnotemark[1]
John Southworth,$^{1}$\footnotemark[1]
\newauthor
Guillermo~Torres,$^{7}$\footnotemark[1]
Guy R Davies,$^{8, 9}$\footnotemark[1]
Warrick Ball,$^{8, 9}$
\newauthor
and William J Chaplin$^{8, 9}$
\\
$^{1}$Astrophysics group, Keele University, Keele, Staordshire, ST5 5BG, UK.\\
$^{2}$Max-Planck-Institut f\"{u}r Sonnensystemforschung, Justus-von-Liebig-Weg
3, 37077, G\"{o}ttingen, Germany.\\
$^{3}$Nicolaus Copernicus Astronomical Center, Polish Academy of Sciences, ul. Rabia{\'n}ska 8, 87-100 Toru{\'n}, Poland. \\
$^{4}$Instituut voor Sterrenkunde, KU Leuven, Celestijnenlaan 200D, B-3001
Leuven, Belgium. \\
$^{5}$Astronomy Department, San Diego State University, 5500 Campanile Drive,
San Diego, CA 92182-1221, USA.\\
$^{6}$Villanova University, Dept. of Astrophysics and Planetary Science, 800
Lancaster Avenue, Villanova PA 19085, USA.\\
$^{7}$Center for Astrophysics, Harvard \& Smithsonian, 60 Garden Street,
Cambridge, MA 02138, USA.\\
$^{8}$School of Physics and Astronomy, University of Birmingham, Birmingham,
B15 2TT, UK.\\
$^{9}$Stellar Astrophysics Centre (SAC), Department of Physics and Astronomy,
Aarhus University, Ny Munkegade 120,\\
DK-8000 Aarhus C, Denmark.
}

\date{Accepted XXX. Received YYY; in original form ZZZ}
\pubyear{2020}

\begin{document}
\label{firstpage}
\pagerange{\pageref{firstpage}--\pageref{lastpage}}
\maketitle

\begin{abstract}
Accurate masses and radii for normal stars derived from observations of
detached eclipsing binary stars are of fundamental importance for testing
stellar models and may be useful for calibrating free parameters in these
model if the masses and radii are sufficiently precise and accurate. We aim to
measure precise masses and radii for the stars in the bright eclipsing binary
AI~Phe, and to quantify the level of systematic error in these estimates. We
use several different methods to model the TESS light curve of AI~Phe combined
with spectroscopic orbits from multiple sources to estimate precisely the
stellar masses and radii together with robust error estimates. We find that
the agreement between different methods for the light curve analysis is very
good but some methods underestimate the errors on the model parameters. The
semi-amplitudes of the spectroscopic orbits derived from spectra obtained with
modern \'echelle spectrographs are consistent to within 0.1\%. The masses of
the stars in AI~Phe are $M_1 = 1.1938 \pm 0.0008\,\rm M_{\odot}$  and $M_2 =
1.2438 \pm 0.0008\,\rm M_{\odot}$, and the radii are \mbox{$R_1 = 1.8050 \pm
0.0022\,\rm R_{\odot}$} and \mbox{$R_2 = 2.9332 \pm 0.0023\,\rm R_{\odot}$}.
We conclude that it is possible to measure accurate masses and radii for stars
in bright eclipsing binary stars to a precision of 0.2\% or better using
photometry from TESS and spectroscopy obtained with modern \'echelle
spectrographs. We provide recommendations for publishing masses and radii of
eclipsing binary stars at this level of precision.
\end{abstract}

%

\section{Introduction}

AI~Phe is a moderately bright (V=8.6) star that was first noted as an
eclipsing binary by \citet{1972IBVS..665....1S}. Photometric monitoring by
\citet{1978IBVS.1419....1R} established that the orbital period is
approximately 24.59 days and that the primary eclipse is total.
\cite{1984ApJ...282..748H} obtained UBVRI light curves of AI~Phe covering both
eclipses and combined their analysis of these light curves with the
spectroscopic orbits for both stars published by  \citet{1979A&AS...36..453I}
to measure the masses and radii of both components. They found that the stars
in AI~Phe are slightly more massive than the Sun, and both have evolved away
from the zero-age main sequence, with the more massive star being a subgiant.
A much-improved spectroscopic orbit for AI~Phe was published by
\citet{1988A&A...196..128A} together with a re-analysis of the UBVRI light
curves by \cite{1984ApJ...282..748H} and new {\it uvby} light curves, and a
spectroscopic estimate of the metal abundance (${\rm [Fe/H]} =-0.14\pm0.1$).
The analysis by Andersen et al. has been the definitive observational study of
AI~Phe until recently and the absolute parameters derived therein have often
been used as a benchmark for testing stellar evolution models of single stars
\citep[e.g.,][]{1988A&A...196..128A, 1997MNRAS.289..869P, 2000MNRAS.318L..55R,
2002A&A...396..551L, 2013ApJ...776...87S, 2015ApJ...812...96G,
2017A&A...608A..62H}. 
   
\citet{2016A&A...591A.124K} used light curves obtained during the WASP
  exoplanet transit survey together with improved spectroscopic orbits for
both components by \citet{2009MNRAS.400..969H} to improve the precision of the
radius measurements for both components of AI~Phe to better than 1\%.
\citet{2018MNRAS.478.1942S} also obtained spectroscopic orbits for both
components of AI~Phe together with spectroscopic observations obtained during
the secondary eclipse which show that the rotation axis of the subgiant
component is not aligned with the orbital axis of the binary system.
\citet{2019A&A...632A..31G} have used the VLTI interferometer to measure the
astrometric orbit of AI~Phe. They used this astrometric orbit combined with
their own high-quality spectroscopic orbits for the two stars to derive their
masses with a precision of better than 0.1\% and a model-independent distance
to the binary of $169.4\pm 0.7$\,pc.
  
The duration of the primary and secondary eclipses of AI~Phe are approximately
15 and 20 hours, respectively. This, combined with the long orbital period of
AI~Phe, makes it difficult to obtain complete light curves from ground-based
observatories. For any mid-latitude observing site it will be necessary to
patch together eclipses observed on different nights, or even from different
observing seasons, to obtain complete coverage of both eclipses.
\citet{2016A&A...591A.124K} found that the orbital period of AI~Phe is not
constant, presumably due to the presence of a third body in the system. The
orbit of the presumed third body is not well characterised, so it is possible
that there is a systematic error in the published radius estimates for this
binary due to inaccuracies in the orbital phases assigned to different parts
of the light curve. This problem of phasing the light curve may explain the
small discrepancies between the values of the orbital eccentricity, $e$,  and
longitude of periastron, $\omega$, obtained in different studies
\citep{2016A&A...591A.124K}.
   
The launch of the Transiting Exoplanet Survey Satellite
\citep[TESS,][]{2015JATIS...1a4003R} afforded us the opportunity to obtain a
high-precision light curve of AI~Phe with complete coverage of both eclipses
observed over a single orbital period of the binary system. Accordingly, we
applied for and were awarded guest-observer status to observe AI~Phe using the
2-minute cadence mode of TESS (G011130, P.I. Maxted; G011083, P.I. He\l
miniak; G011154, P.I. Pr\v{s}a). TESS is designed to produce very high quality
photometry of bright stars (100 ppm per hour or better for stars with I$_{\rm
c}\approx 8$) and AI~Phe shows very well-defined total eclipses, so we expect
a very small random error on the parameters derived from the analysis of the
TESS light curve. Light curves for eclipsing binary stars of this quality have
only become available within the past few years so it is not yet clear to what
extent the parameters derived are affected by systematic errors due to the
methods used for the analysis. In order to get a better understanding of these
systematic errors we present here several independent analyses of the TESS
light curve of AI~Phe conducted by researchers using a variety of different
software packages that they have previously used to analyse light curves of
eclipsing binary stars. We also compile and compare the published
spectroscopic orbits for AI~Phe, both with each other and, where possible, to
parameters derived from the light curve analysis. These results are then
combined to obtain accurate mass and radius estimates for both components in
AI~Phe to a precision of better than 0.2\%.
 
\begin{figure}
\centering
\includegraphics[width=0.47\textwidth]{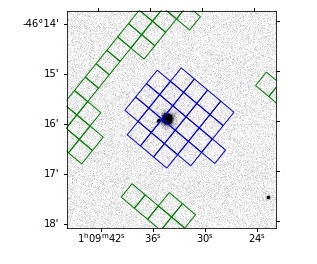}
\caption{Skymapper r$^{\prime}$-band image of AI~Phe overlaid with the
aperture used to calculate the TESS light curve of AI~Phe. Pixels used to
measure the flux are plotted in blue and pixels used to calculate the
background value are plotted in green. The Skymapper data are displayed as an
inverse logarithmic gray-scale image. }
\label{TESS_Aperture_Overlay_Skymapper}
\end{figure}
%

 \section{TESS light curve analysis}
The sky region around AI~Phe is shown in
Fig.~\ref{TESS_Aperture_Overlay_Skymapper} overlaid with the location of the
pixels in the TESS images that are used to calculate the TESS light curve. The
faint star to the east of AI~Phe that is included in the TESS photometric
aperture is listed in the TESS input catalogue (TIC) as TIC~616203794 with an
estimated magnitude in the TESS band of T=13.67. The estimated magnitude of
AI~Phe listed in the TIC is T=7.96, so this faint companion contributes only
about 0.5\% to the total flux measured by TESS. The pixels used to estimate
the background level in the TESS images do not contain any stars brighter than
T=13.5.  
 
\begin{figure}
\centering
\includegraphics[width=0.47\textwidth]{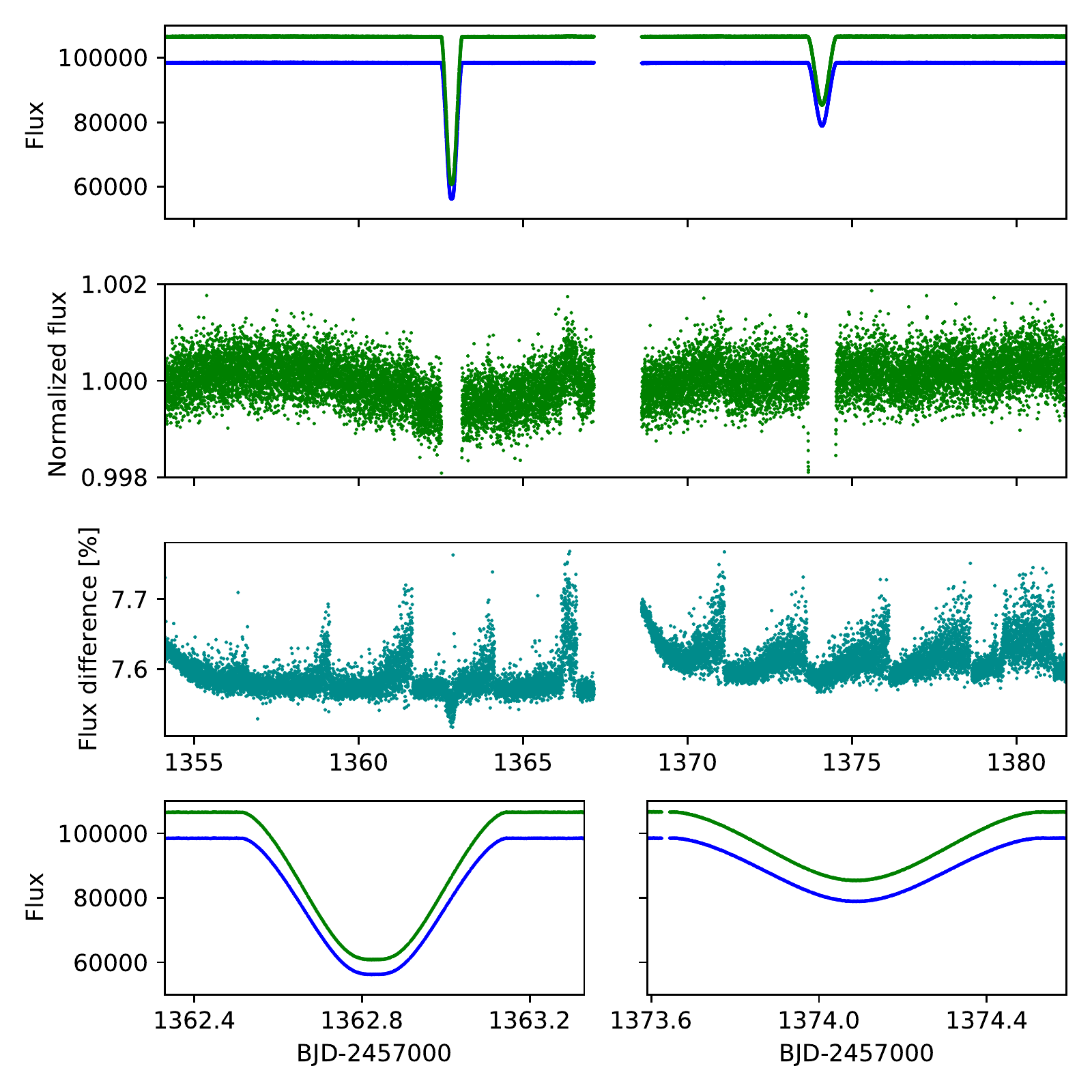}
\caption{The TESS light curve of AI~Phe. The lower (blue) curves show the {\tt
SAP\_FLUX} values and the upper (green) curves shows the {\tt PDCSAP\_FLUX}
values from the MAST archive data file. Only the {\tt PDCSAP\_FLUX} values are
shown in the second panel down from the top. The flux difference value shown
in the third panel down are $1-{\tt SAP\_FLUX/PDCSAP\_FLUX}$. Only data with a
QUALITY flag value of 0 are shown in this figure.}
\label{TESSLightCurve}
\end{figure}

TESS observations of AI~Phe took place from 2018-08-22 to 2018-09-20
(Sector~2). The TESS light curve files downloaded from the Mikulski Archive
for Space Telescopes (MAST) contain two measurements of the flux labelled {\tt
SAP\_FLUX} and {\tt PDCSAP\_FLUX}. {\tt SAP\_FLUX} light curves are the result
of simple aperture photometry, i.e., the total counts measured within the
photometric aperture from the calibrated TESS images corrected for the
estimated background flux. {\tt PDCSAP\_FLUX} light curves include a
correction to remove instrumental systematic variations by fitting and
removing those signals that are common to all stars on the same detector. The
{\tt SAP\_FLUX} and {\tt PDCSAP\_FLUX} light curves for AI~Phe are shown in
Fig.~\ref{TESSLightCurve}. The light curve files also include a QUALITY flag
for each measurement.  In this study we only use data with QUALITY=0, i.e.,
for which there are no known issues with the measurement. Of the 19\,737 flux
measurements for AI~Phe in the TESS light curve file, 18\,303 (93\%) have
QUALITY=0. The meta-data provided with the MAST light curve file includes a
parameter {\tt CROWDSAP} that gives the ratio of the target flux to the total
flux in the photometric aperture. For the AI~Phe data this has the value
0.99993235, so we assume that the contamination of the aperture by
TIC~616203794 has not been accounted for in the creation of the {\tt
PDCSAP\_FLUX} light curve.

In the following subsections we describe in detail the various binary star
models, optimisation methods and parameter error estimation techniques that we
have used to analyse the TESS light curve of AI~Phe. The methods are
summarised in Table~\ref{models}. In all of the following descriptions, star 1
(``primary star'') is the F7\,V star that is totally eclipsed at phase 0
(``primary eclipse'') by star 2 (``secondary star'') the K0\,IV star. We use
the usual notation $r_i=R_i/a$ where $r_i$ is the ``fractional radius'' of
star $i$,  $R_i$ is the actual radius of star $i$ and $a$ is the semi-major
axis of the binary orbit.   The TESS light curve of AI~Phe obtained during
TESS Cycle 1 contains only one eclipse of each type so it is not possible to
determine the orbital period from these observations. For all of the light
curve analysis methods described below we fixed the orbital period at the
nominal value P=24.5924\,d, close to the value obtained by
\cite{2016A&A...591A.124K} from WASP observations of AI~Phe obtained from 2006
to 2011.
   
\begin{figure}
\centering
\includegraphics[width=0.47\textwidth]{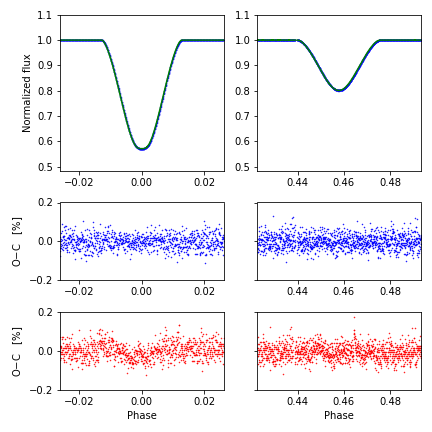}
\caption{Upper panel: Best-fit model from Run A (green line) for the TESS light curve of AI~Phe (blue points). Middle panel: residuals from the best-fit model for Run A. Lower panel: residuals from the best-fit model for Run E. }
\label{MaxtedLCFit}
\end{figure}

   \begin{figure*}
   \centering
   \includegraphics[width=0.98\textwidth]{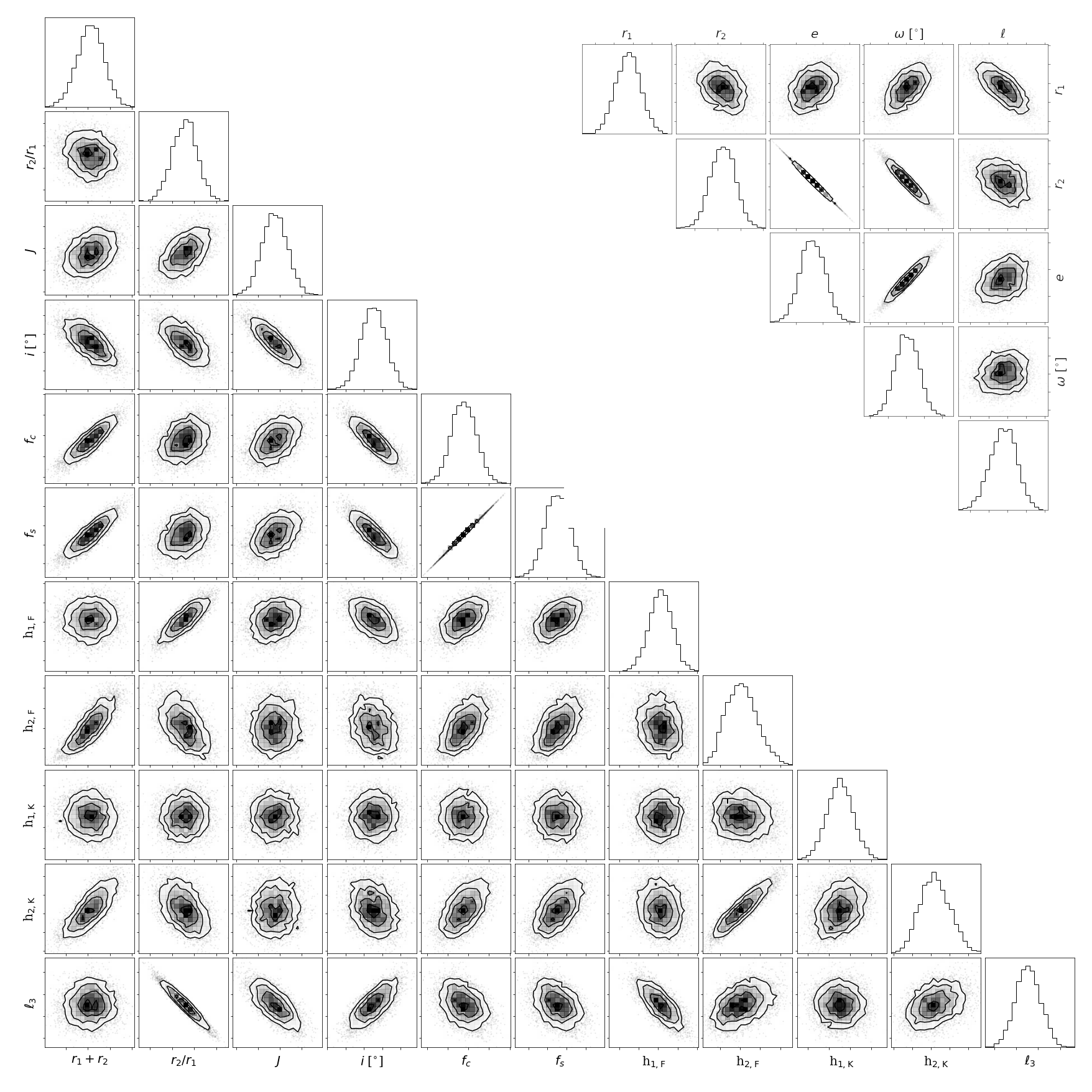}
      \caption{Parameter correlation plots for selected parameters from Run A. Model parameters not shown in this corner plot ($f$ and $\sigma_f$) do not show any strong correlation with the other model parameters shown here. Parameter correlations are shown in the upper-right corner for quantities derived from the model parameters, including $\ell = S_{\rm T}\times k^2 = 1.3212 \pm 0.0054$, the flux ratio in the TESS band. }
         \label{corner_maxted}
   \end{figure*}
%

   \begin{figure*}
   \centering
   \includegraphics[width=0.75\textwidth]{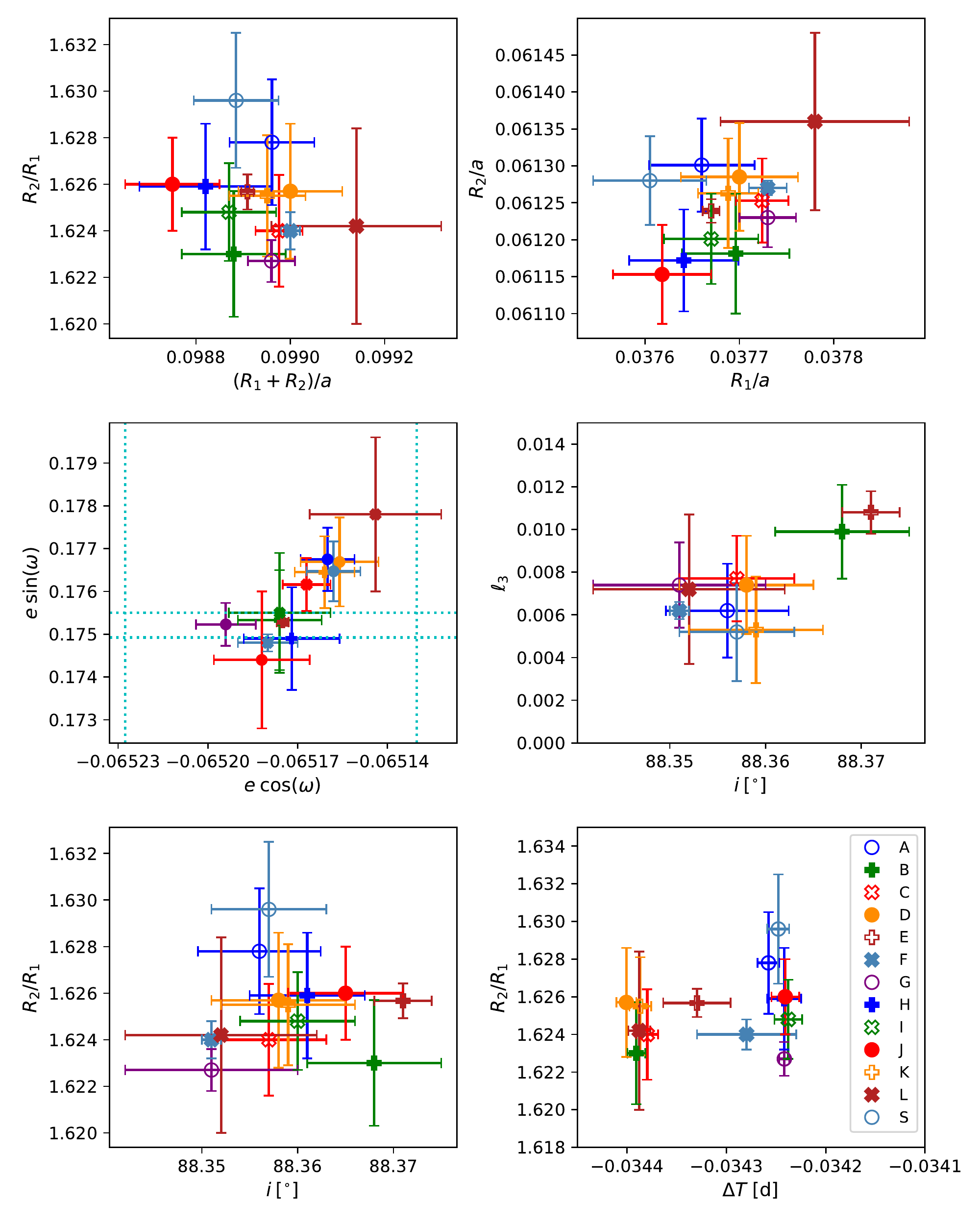}
      \caption{Parameters of AI~Phe obtained from the TESS light curve using different light curve models. The dashed lines in the plot of $e\sin\omega$ versus $e\cos\omega$ are the values determined from the spectroscopic orbits of  \citet{2018MNRAS.478.1942S} and  \citet{2019A&A...632A..31G}.  $\Delta T$  is the measured time of mid-eclipse relative to BJD$_{\rm TDB}=  2458362.86273$, the time of mid-eclipse predicted using the ephemeris from \citet{2018A&A...620C...5K}.}
         \label{ResultsLCFit}
   \end{figure*}
%

\begin{table*}
\caption[]{Summary of models and optimisation methods. }
\label{models}
\begin{tabular}{lllllll}
\hline
\noalign{\smallskip}
Run & Investigator &Model &  Optimization & Limb-darkening & Detrending & Notes \\
\noalign{\smallskip}
\hline
\noalign{\smallskip}
A & Maxted & {\tt ellc} &  emcee & power-2 & celerite &  \\ 
B & He{\l}miniak & \textsc{jktebop}   & L-M & quadratic & sine+poly & Monte Carlo error estimates. \\  
C & Torres &  EB &  emcee & quadratic & spline &  Quadratic l.d. coeffs. fixed.\\ 
D & \texttt{"}  &  \texttt{"}  &  \texttt{"}  & \texttt{"}  & \texttt{"}  &  \\ 
E & Graczyk &  WD2007 & L-M &logarithmic      &-- & Fixed l.d. coefficients \\ 
F & Johnston & phoebe 1.0 & emcee & square-root & -- &  \\ 
G & Pr\v{s}a & phoebe 2.1 & MCMC & grid & legendre &  \\
H & Orosz & ELC & DE-MCMC & logarithmic & polynomial & \\
I & Orosz & \texttt{"} & \texttt{"} & square-root & \texttt{"} & \\
J & Orosz & \texttt{"} & \texttt{"} & quadratic   & \texttt{"} & \\
K & Southworth & \textsc{jktebop} & L-M & quadratic & polynomial & \\
L & Southworth & \textsc{jktebop} & L-M & cubic & polynomial & \\
S & Maxted & {\tt ellc} &  emcee & power-2 & celerite &  Same as Run A with {\tt SAP\_FLUX}\\
\noalign{\smallskip}
\hline
\end{tabular}   
\end{table*}

\subsection{Run A -- Maxted, {\tt ellc}}
  For Run A we used version 1.8.5 of the binary star model {\tt ellc}\footnote{\url{https://github.com/pmaxted/ellc}} \citep{2016A&A...591A.111M} to fit the {\tt PDCSAP\_FLUX} light curve. We used the option available in this version of {\tt ellc} to use the power-2 limb-darkening law \citep{1997A&A...327..199H, 2018A&A...616A..39M} for both stars. The trends in the data were first removed by fitting a simple Gaussian process (GP) model to the data between the eclipses in 1-hour bins using the {\tt celerite} software package \citep{2017AJ....154..220F} and then dividing the measured fluxes by the best-fit GP model evaluated at all times of observation. 
  
  The free parameters in the fit are:  a flux scaling factor, $f$; the sum of the fractional radii, $r_{\rm sum} = r_1+r_2 = (R_1+R_2)/a$; the ratio of the radii, $k = R_2/R_1$; the surface brightness ratio averaged over the stellar disks in the TESS band, $S_{\rm T}$; the orbital inclination, $i$; the time of primary eclipse, T$_0$;  $f_s = \sqrt{e}\sin(\omega)$ and $f_c = \sqrt{e}\cos(\omega)$; ``third light'', $\ell_3$; the parameters of the power-2 limb-darkening law for star~1, $h_{\rm 1, F}$ and $h_{\rm 2, F}$; the parameters of the power-2 limb-darkening law for star~2, $h_{\rm 1, K}$ and $h_{\rm 2, K}$;  the standard error per observation, $\sigma_f$.

 We used Table 2 from \citet{2018A&A...616A..39M} to estimate the following limb darkening parameters -- $h_{\rm 1, F}=0.818$, $h_{\rm 2, F}=0.420$, $h_{\rm 1, K}=0.770$, $h_{\rm 2, K}=0.424$. We used Gaussian priors centred on these values with standard deviation of 0.05 and 0.25 for the $h_1$ and $h_2$ parameters, respectively. Based on the results in \citet{2018A&A...616A..39M} we expect the tabulated values to be much more accurate than these rather generous priors. The light travel time across the orbit was included in the model assuming $a=47.868\,{\rm R}_{\odot}$ \citep{2015MNRAS.448..946B}. We used {\sc emcee} \citep{2013PASP..125..306F}, a {\sc python} implementation
of an affine invariant Markov chain Monte Carlo (MCMC) ensemble sampler, to
calculate the posterior probability distribution (PPD) of the model parameters. In order to speed up the calculation we assumed that the stars are spherical and used the ``default'' grid size for both stars. Another speed-up we used is to only calculate 1-in-4 observations using the {\tt ellc} model and then to use linear interpolation to calculate intervening observations. To estimate the numerical noise in our light curves we compared a light curve computed with these settings to one computed with the ``very-fine'' grid size and evaluated at every observed point. The RMS residual of the difference through the eclipses is 33\,ppm, about 10 times less than the standard error per observation.  The oblateness of both stars is very small ($< 0.08\%$) so the use of spheres to approximate their shapes is very reasonable in this case.

 The PPD was sampled using 1000 steps and 256 walkers following a ``burn-in'' phase of 1000 steps. The light curve was analysed in two sections, each centred on an eclipse and with a width approximately twice the eclipse width. The convergence of the chain was evaluated by-eye from plots of parameters values versus step number. The typical auto-correlation length of the chains is about 70 steps so we thinned the chains by 100 steps before calculating the mean and standard deviation of the parameters given in Tables~\ref{geompar} and \ref{radpar}. The best-fit model light curve fit to  the data used in the {\sc emcee} analysis is shown in Fig.~\ref{MaxtedLCFit}

\subsection{Run B -- He{\l}miniak, \textsc{jktebop}}
 The analysis for Run B was done using {\sc \textsc{jktebop}}\footnote{{\sc \textsc{jktebop}} is written in {\sc fortran77} and the source code is available at  {\tt http://www.astro.keele.ac.uk/jkt/codes/jktebop.html}} version 34  \citep[][and references therein]{2013A&A...557A.119S}. This code uses the  Levenberg-Marquardt algorithm \citep{1992nrfa.book.....P} to optimise the  parameters of the {\sc ebop} light curve model \citep{1981psbs.conf..111E,1981AJ.....86..102P} from a least-squares fit to a light curve ({\tt PDCSAP\_FLUX}) expressed in magnitudes. We used the quadratic limb-darkening law option for both stars. The integration ring size used for numerical integration of the observed fluxes was 1 degree. This is sufficient to reduce numerical noise to less than 10 ppm.

The free parameters in the least-squares fit for Run B are $J$ (the central surface brightness ratio), $k$, $r_{\rm sum}$, $i$, $e$, $\omega$, $\ell_3$, T$_0$, $m_0$ (magnitude zero-point), and the coefficients of the quadratic limb darkening law for both stars. The parameter $J$ is defined as the ratio of the surface brightness at the centre of the stellar discs, rather than the disk-average value $S_{\rm T}$ used for Run A. The residuals from an initial fit to the data were analysed using a Lomb-Scargle periodogram \citep{1992nrfa.book.....P}. From this periodogram two significant frequencies corresponding to periods near 26\,d and 2.4\,d were identified. The 2.4-day periodicity is due to "momentum dump" events when the space craft reaction wheel speeds are periodically reset to low values. This can be seen in the difference between the {\tt PDCSAP\_FLUX} and {\tt SAPFLUX} values shown in Fig.~\ref{TESSLightCurve}.

We include sine waves with these two frequencies in the least-squares fit to the data together with a 3$^{\rm rd}$ degree polynomial to model an additional trend seen in the residuals near the start of the data set.  {\sc \textsc{jktebop}} includes options to modulate the flux of either star using these sine waves and polynomials, or to simply apply these trends as an overall scaling factor. We found that there is very little difference in the results for different combinations of these options so here we only quote the results for applying them as an overall scaling factor. We kept the frequency of the sine waves fixed at the values determined from the periodogram but their phases and amplitudes were included as free parameters in the fit, together with the coefficients of the 3$^{\rm rd}$ degree polynomial.

All data were equally weighted in the least-squares fit and the errors on the parameters were estimated using a Monte Carlo simulation assuming Gaussian uncorrelated noise with the same standard deviation as the rms residual from the best fit. We checked these error estimates using the prayer-bead Monte Carlo method, which tended to give slightly lower error estimates. The parameters of the sine and polynomial de-trending functions were kept fixed at their best-fit values during these Monte Carlo simulations.

\subsection{Runs C and D -- Torres, EB}
 For Runs C and D the light curve analysis used the {\tt PDCSAP\_FLUX} photometry, and was done with the EB program of \citet{2011ApJ...742..123I}, an improved version of the original {\sc ebop} light curve model, combined with {\sc emcee} to explore the model parameter space. With the default integration ring size in EB the contribution of numerical noise to the computed light curves is $\approx $30\,ppm or less. Trends in the data were removed by dividing thorough by a spline fit by least squares to the data between the eclipses, so computation of the ellipsoidal and reflection effects in the light curve were disabled.  The free parameters in both fits were $J$, $k$, $r_{\rm sum}$, $\cos i$, $f_s = \sqrt{e}\sin(\omega)$, $f_c = \sqrt{e}\cos(\omega)$,  $\ell_3$, T$_0$ and $\ln f_{\sigma}$, where $f_{\sigma}$ is a scaling factor for the errors on each measurement.
 
 For Run C we used a quadratic limb darkening law for both stars with the linear coefficients as free parameters and the quadratic coefficients fixed at the values for the TESS band of 0.3107 and 0.2162 for stars 1 and 2, respectively \citep{2011A&A...529A..75C}. For Run D we used the transformed limb-darkening parameters $q_1$ and $q_2$ defined by \citet{2016MNRAS.455.1680K} to efficiently sample the dependence of the posterior probability distribution (PPD) on both quadratic limb-darkening coefficients for both stars. Light travel time within the binary was accounted for in the analysis, adopting the value 100.27\,km\,s$^{-1}$ for the sum of the radial-velocity semiamplitudes, from \citet{2009MNRAS.400..969H}. The sampling of the PPD was done using 100 walkers and 40\,000 steps following a burn-in phase with 35\,000 steps. The results shown in Tables~\ref{geompar} and \ref{radpar} were calculated from the mode of the posterior distributions and the 1-sigma errors based on half the range between the 16$^{\rm th}$ and 84$^{\rm th}$ percentiles.

\subsection{Run E -- Graczyk, WD2007}
 For Run E we used the Wilson-Devinney light curve model version WD2007 \citep{2007ApJ...661.1129V} with  8392 grid points for both stars and logarithmic limb-darkening laws for both stars. We used the Cousins $I_{\rm C}$ band as an approximation to the TESS photometric pass band and set the secondary star effective temperature to 5086\,K. The free parameters in the least-squares fit are the effective temperature of the primary star, the values of the potentials defining the size and shapes of each star, the orbital inclination, $e$ and $\omega$, $\ell_3$, a flux-scaling factor and a phase shift relative to the nominal time of mid-eclipse. The mass ratio used to calculate the Roche potential was $q=M_2/M_1 = 1.04017$. The data were normalised using a spline fit to {\tt PDCSAP\_FLUX} values between the eclipses. Only data in the region of the eclipses (2274 observations) were included in the analysis. The best fit model parameters were found by repeated application of the differential corrections calculated by WD2007. To estimate the numerical noise in our final model light curve we compared it to a light curve computed with the same model parameters but a much finer grid spacing. In the regions of the eclipse the rms difference between these two light curves is 140\,ppm.
 
 The radii quoted here are the arithmetic means of the radii at the side, pole and substellar point of the Roche equipotential surfaces. A Monte Carlo calculation was used to calculate the standard errors for quantities derived from the free parameters of the model assuming that their joint probability distribution is a multivariate Gaussian distribution. The covariance matrix for this multivariate Gaussian distribution was derived from the correlation matrix and parameter standard deviations provided in the output from WD2007. Some small systematic trends ($\sim 300$\,ppm) are seen in the residuals from the fit to primary eclipse (Fig.~\ref{MaxtedLCFit}). This is likely to be a consequence of using Cousins I$_{\rm C}$ band as an approximation to the TESS band for both the interpolation the limb-darkening coefficients and for the calculation of the flux ratio from the effective temperatures of both stars.

\subsection{Run F -- Johnston, Phoebe 1.0}
Phoebe 1.0 is an enhanced version of the widely-used Wilson-Devinney (WD) light curve model. It incorporates simple reflection, parameterised limb-darkening, and gravity darkening, amongst other physics \citep{2005ApJ...628..426P}. We used a square-root limb-darkening law with coefficients for the TESS pass band interpolated from a pre-calculated grid of coefficients  for each model. We increased the fine-grid resolution for the PHOEBE model by a factor three compared to the standard grid size in the original WD model in order to better reproduce the points of ingress and egress. We adopted the solution of \citet{2016A&A...591A.124K} as our starting solution and determined errors using the ensemble MCMC sampling code {\sc emcee} \citep{2013PASP..125..306F}. We fixed the gravity darkening coefficients, period, and primary effective temperature to the value from \citeauthor{2016A&A...591A.124K}. To estimate the numerical noise in our final model light curve we compared it to a light curve computed with the same model parameters but a much finer grid spacing. In the regions of the eclipse the rms difference between these two light curves is 126\,ppm.

We did not include radial velocities in our fit but, in order to propagate forward any influence of a varying mass ratio, we applied a Gaussian prior on this parameter according to the value by Kirkby-Kent et al. (2016). Additionally, we sampled the reference time of mid-eclipse $T_0$, $\cos i$,  $e\cos\omega$, $e\sin\omega$, the secondary effective temperature, the primary and secondary potentials and luminosities, and the third light. We fitted the {\tt PDCSAP\_FLUX} lightcurve provided by MAST with no further detrending. We used 128 chains and ran the algorithm for 10,000 iterations. We ensured solution convergence by running the sampler for at least five times the autocorrelation time of all parameter chains. We discarded all iterations before convergence as part of the burn-in phase. 

\subsection{Run G -- Pr\v{s}a, PHOEBE 2.1}
PHOEBE 2 \citep{phoebe2} is a redesigned modelling suite for computing observable stellar properties that extends the previous version by adding missing physics and by increasing modelling fidelity. All technical and computational considerations are described in \citet{prsa2018}, including tests that were done to demonstrate that there is very little numerical noise in the light curves computed with this model.

The light curve of AI~Phe was detrended by fitting a chain of 5th order Legendre polynomials connected at data gaps to the sigma-clipped out-of-eclipse regions. PHOEBE 2.1 was used to build a model. Direction-dependent emergent intensities for each surface element were computed from \citet{ck2004} model atmospheres; limb darkening was interpolated on a grid of 72 $\mu$ vertices from \citeauthor{ck2004} atmospheres synthesized by the \texttt{spectrum} code \citep{1999ascl.soft10002G}; reflection/irradiation included both heating and scattering components explained in \citet{2018ApJS..237...26H}. The eccentricity and argument of periastron were replaced with their orthogonalized components, $e \cos \omega$ and $e \sin \omega$. The adjusted parameters include the time of superior conjunction, orbital inclination, effective temperature ratio, passband luminosity, third light contribution in the TESS passband, orthogonalized eccentricity and argument of periastron, and the sum and ratio of stellar radii. Initial parameter optimization was done using Nelder and Mead's Simplex method \citep{2005ESASP.576..611P}, and solution sampling was done using Markov Chain Monte Carlo sampler {\sc emcee} \citep{2013PASP..125..306F}. Convergence was assessed by the Gelman-Rubin criterion \citep{GR92}. In total, 128 walkers were used over a little over $10^5$ iterations. Integrated surface brightness and flux ratios are computed at phase 0.25.

\subsection{Runs H, I and J -- Orosz, ELC}
 Runs H, I and J are identical except that a different limb-darkening law was used in each case, as listed in Table~\ref{models}. The light curve was analysed in two sections, each centred on an eclipse and with a width approximately twice the eclipse width. A fifth-order polynomial fit by least-squares to the data either side of each eclipse was used to divide-through all the data in each section so that the mean level out-of-eclipse was 1. The final adopted light curve has $N = 2193$ measurements with a median uncertainty per observation of 348\,ppm. We used the ELC code \citep{2000A&A...364..265O} to model the light curve. ELC has been modified to use the technique of \cite{2018AJ....156..297S} for the fast computation of eclipses of spherical stars. The tolerance settings in the numerical integrator used in ELC are set such that numerical noise in these light curves is well below 1\,ppm. We used the optimiser available in ELC based on the DE-MCMC algorithm \citep{ter2006markov} for the results presented here. Similar runs using the optimiser based on the nested sampling algorithm \citep{skilling2006nested} gave similar results but with much smaller errors on the parameters. The primary temperature was fixed at $T_1$ = 6310\,K. The free parameters in the fit were the inclination $i$, the time of primary eclipse $T_0$, the temperature ratio $T_2/T_1$, the sum of the fractional radii $r_1 + r_2$, the difference of the fractional radii $r_1-r_2$, $e\,\cos\omega$,$e\,\sin\omega$, $\ell_3$, and the four limb-darkening coefficients, two for each star. The temperature ratio is used to calculate the central surface brightness ratio, $J$,  by integration of a model spectral energy distribution for each star over the TESS passband. For the analysis of a single light curve, this is equivalent to including $J$ as a free parameter. 
 
 For each run, 50 chains were used and the chains were allowed to evolve for 100\,000 generations. The chains were allowed to ``burn in'' for 1000 generations. Thereafter, the posterior sample was constructed by using every thousandth generation.  The uniform priors imposed on the model parameters generally span a much larger range than the width of posterior distribution. The exception is third-light for which we found the posterior was approximately a uniform sampling of the prior $\ell_3 = [0,0.05]$.

\subsection{Runs K and L: Southworth and \textsc{jktebop}}

For these runs we used version 38 of the \textsc{jktebop} code, which differs only in small ways from version 34 used for Run B. We present results for the cases where limb-darkening was implemented using the quadratic law (Run K) and the cubic law ($I(\mu) = 1 - u(1-\mu) - v(1-\mu)^3$;  Run L) --  the two runs are the same in all other respects. We also tried square-root limb-darkening, finding that the results were very similar but the coefficients were highly correlated; logarithmic limb-darkening, for which the correlations were so high the fit failed to converge quickly; and linear limb-darkening, which we found to be inadequate for the current data. See \citet{Me08mn} for the definitions of the limb-darkening laws used.

We obtained the TESS light curve of AI\,Phe and converted it to magnitude units. We rejected the data more than twice the eclipse duration away from an eclipse midpoint, in order to speed up the fits to the data.  We fitted for the sum and ratio of the fractional radii ($r_{\rm sum}$ and $k$), the orbital inclination ($i$), the central surface brightness ratio ($J$), the strength of the reflection effects for the stars, and the time of the midpoint of the primary eclipse ($T_0$). Third light ($\ell_3$) was also included as a fitted parameter because of the presence of the fainter nearby star: we found a small value for $\ell_3$ that was nevertheless significant at the 2$\sigma$ level. All fits included two first-order polynomials to model the out-of-eclipse brightness of the system, one for each eclipse, as de-trending functions.

Theoretical values for the quadratic limb-darkening coefficients of the two stars in the TESS passband were obtained from \citet{Claret17aa}. Cubic limb-darkening coefficients are not available from this source so were obtained by fitting the quadratic trend with the cubic law. Initial fits for both quadratic and cubic limb darkening  showed good results when the limb-darkening coefficients were fixed at these theoretical values. For the quadratic law we obtained a significant improvement when fitting for the linear coefficient for each star ($\Delta\chi^2 = 220$), but no further improvement when fitting for both coefficients for each star ($\Delta\chi^2 = 0.2$). For the cubic law the values are $\Delta\chi^2 = 228$ and $5.5$, respectively. In both cases all limb-darkening  coefficients were physically realistic and measured with reasonable precision. We conclude that the theoretical coefficients are clearly worse than the fitted values, and that it is necessary to fit for at least one limb-darkening  coefficient for each star. Our final results are given for the cases where the two linear limb-darkening  coefficients are fitted using the quadratic law (Run K) and all four limb-darkening  coefficients are fitted using the cubic law (Run L).

Fits were made to the data using the Levenberg-Marquardt method \citep{1992nrfa.book.....P}. We tried values of the integration ring size from 0.2 to 5 degrees. This change had a negligible effect on the results so we used a size of 5 degrees for the final fits. The formal uncertainties from the covariance matrix often underestimate the true uncertainties, so we obtained parameter uncertainties in two other ways. First, we used 1000 Monte Carlo simulations \citep{Me++04mn2}. Second, we used the residual-permutation algorithm implemented in \textsc{jktebop} \citep{Me08mn}. The final error bar for each parameter is the larger of the two options for that parameter, and in most cases this comes from the residual-permutation simulations.

The final results for Runs K and L are shown in Tables \ref{geompar} and \ref{radpar}. The differences between the best-fitting values for the two runs are due only to the treatment of limb-darkening. The larger error bars for Run L versus Run K are because of the inclusion of four rather than two limb-darkening coefficients as fitted parameters.

\subsection{Run S -- Maxted, {\tt ellc}}
According to the TESS archive manual, at times the fitting method used to remove instrumental trends from the {\tt SAP\_FLUX} measurements is known to remove true astrophysical signals from the {\tt PDCSAP\_FLUX} values. In order to better understand if this is a problem for the TESS light curve of AI~Phe, we repeated Run A using {\tt SAP\_FLUX} measurements instead of {\tt PDCSAP\_FLUX}.

\begin{table*}
\caption[]{Geometric and orbital parameters derived from the analysis of the TESS light curve of AI~Phe. The values in the final row are the mean and sample standard deviation for all runs. The time of primary eclipse, $T_0$ is given as BJD$_{\rm TDB}-2458362$.}
\label{geompar}
\begin{center}
  \begin{tabular}{@{}lllllllll}
\hline
\noalign{\smallskip}
 \multicolumn{1}{@{}l}{Run} &
 \multicolumn{1}{l}{$r_1=R_1/a$} &
 \multicolumn{1}{l}{$r_2 = R_2/a$} &
 \multicolumn{1}{l}{$r_1+r_2$} &
 \multicolumn{1}{l}{$k=r_2/r_1$} &
 \multicolumn{1}{l}{$i$ [$^{\circ}$] }&
 \multicolumn{1}{l}{$e\cos\omega$} &
 \multicolumn{1}{l}{$e\sin\omega$} &
 \multicolumn{1}{l}{$T_0$}
 \\
\noalign{\smallskip}
\hline
\noalign{\smallskip}
A & 0.037660(56) & 0.061301(63) &  0.09896(9) & 1.6278(27) & 88.356(6) & $ -0.065160(9)$ & 0.1768(7)  & 0.82847(1) \\
B & 0.037696(57) & 0.061181(81) & 0.09888(11) & 1.6230(27) & 88.368(7) & $-0.065176(14)$ & 0.175(1)   & 0.82834(1) \\
C & 0.037724(28) & 0.061253(57) &  0.09898(5) & 1.6240(24) & 88.357(6) & $ -0.065167(8)$ & 0.1762(6)  & 0.82835(1) \\
D & 0.037700(62) & 0.061285(73) &  0.09900(11)& 1.6257(29) & 88.358(7) & $ -0.065156(13)$& 0.177(1)   & 0.82833(1) \\
E & 0.037671(09) & 0.061239(16) &  0.09891(1) & 1.6256(8)  & 88.359(3) & $-0.065175(2)$  & 0.1753(1)  & 0.82840(3) \\
F &   0.03773(2) &   0.06127(1) &  0.09900(2) &  1.6240(8) & 88.351(1) & $  -0.06518(1)$ & 0.1748(2)  & 0.82845(5) \\
G &   0.03773(3) &   0.06123(4) &  0.09896(5) &  1.6227(9) & 88.351(9) & $ -0.065194(1)$ & 0.17523(5) & 0.82849(1) \\
H & 0.037641(58) & 0.061172(69) & 0.09882(14) & 1.6259(27) & 88.361(6) & $-0.065172(16)$ & 0.1749(12) & 0.82849(2) \\
I & 0.037670(50) & 0.061201(61) & 0.09887(10) & 1.6248(21) & 88.360(6) & $-0.065176(17)$ & 0.1755(14) & 0.82849(1) \\
J & 0.037618(52) & 0.061153(67) & 0.09875(10) & 1.6260(20) & 88.365(6) & $-0.065182(16)$ & 0.1744(16) & 0.82849(1) \\
K & 0.037688(32) & 0.061263(74) & 0.09895(8)  & 1.6255(26) & 88.359(7) & $-$0.065161(10) & 0.1765(8)  & 0.82834(1) \\
L & 0.03778(10)  & 0.06136(12)  & 0.09914(18) & 1.6242(42) & 88.352(10)& $-$0.065144(22) & 0.1778(18) & 0.82834(1) \\
S & 0.037605(60) & 0.061280(60) &  0.09889(9) & 1.6296(29) & 88.357(6) & $ -0.065158(9)$ & 0.1765(7)  & 0.82848(1) \\
\noalign{\smallskip}
\hline
\noalign{\smallskip}
--& 0.037686(49) & 0.061242(60) &  0.09893(10) & 1.6252(19) & 88.359(6) & $ -0.065170(14)$ & 0.1758(1) & 0.82842(7) \\
\noalign{\smallskip}
\hline
\end{tabular}
\end{center}
\end{table*}
%
 
   \begin{figure}
   \centering
   \includegraphics[width=0.47\textwidth]{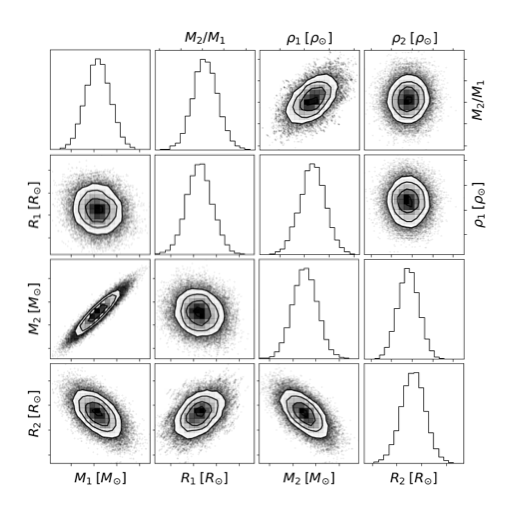}
      \caption{Parameter correlation plots for the mass, radius and density of the stars in AI~Phe.}
         \label{MassRadiusPlot}
   \end{figure}
%

\section{Asteroseismic non-detection}

We have searched the TESS light curves for signatures of solar-like oscillations from both components.  We used the same analysis as described in \citet{2016AN....337..774D} to search for the characteristic pattern of modes of oscillation in the periodogram of the TESS light curve after dividing through the model of the eclisping binary from Run A.  All searches returned posterior distributions consistent with the prior distribution which we use as evidence for a non-detection of modes of oscillation, i.e., the TESS data provides no constraint on the asteroseismic properties of AI~Phe.  

\begin{figure}
   \centering
   \includegraphics[width=0.47\textwidth]{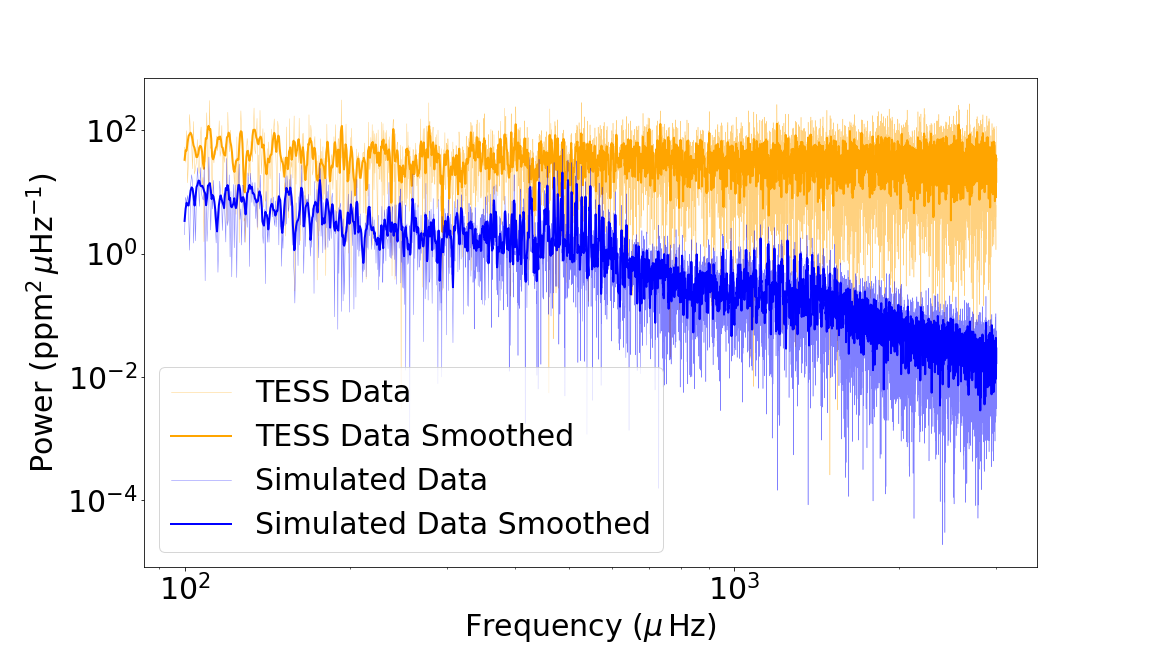}
      \caption{TESS periodogram and simulated noiseless asteroseismic periodogram for AI~Phe.}
         \label{noseismic}
   \end{figure}

An asteroseismic non-detection at this apparent magnitude for these two stars is consistent with predictions \citep[i.e.,][]{2019ApJS..241...12S}.  Figure \ref{noseismic} shows the TESS periodogram together a simulated asteroseismic data set generated following \citet{2018ApJS..239...34B}.  In order to find stellar properties to input into the asteroseismic simulation we used MESA \citep[r10398,][]{Paxton2018}. We optimised to match the observed properties by varying the common age, common initial chemical composition, and individual mixing length parameters.  While the details of the periodogram in Figure \ref{noseismic} will be sensitive to the above procedure the broad large-scale details are insensitive to the choices made above, and at the level of detail displayed, are expected to be accurate. 

The simulated data, in the absence of noise, clearly show two broad humps of modes at around $~ 500$ and $~ 1050 \, \mathrm{\mu Hz}$.  While the highest frequency hump associated with the primary star is substantially lower than the noise floor, the modes of the secondary star are predicted to be close to the TESS AI~Phe noise levels.  A robust asteroseismic detection of the secondary would require a sizeable reduction in the noise levels and/or an extended temporal baseline for the observational time series.  

Despite the non-detection here, AI~Phe remains an excellent target for testing or calibrating asteroseismology, especially given the very high-precision observations given above.  The PLATO mission \citep{2014ExA....38..249R} will provide a larger effective aperture compared with TESS and as such will have significantly reduced noise levels.  At a level of $200$ ppm per cadence, which is indicative of what we might expect with PLATO, we would expect to detect high-quality asteroseismic signals from both components of AI~Phe.   

%

\section{Masses and radii}

 \citet{2009MNRAS.400..969H}, \citet{2018MNRAS.478.1942S} and  \citet{2019A&A...632A..31G} have all published spectroscopic orbits for both stars based on radial velocities (RVs) measured from spectra with good signal-to-noise (S/N$\ga 30$) obtained on \'echelle spectrographs with a resolving power $R\ga 60\,000$. The spectroscopic orbit by \citet{2019A&A...632A..31G} was determined from a simultaneous fit to 33 RVs obtained with the HARPS spectrograph together with the astrometric orbit fit to measurements of the relative positions of the two stars obtained with the VLTI interferometer. The HARPS RVs were first corrected for the motion of the binary system due to a third body in the system with a long, eccentric orbit around AI~Phe ($P_3 \ga 100$\,y, $e\sim 0.8$). As \citet{2019A&A...632A..31G} note, the agreement between the parameters derived from these three independent studies is extraordinarily good. We used the following weighted-mean values for the semi-amplitudes to calculate the masses and radii of AI~Phe: $K_1 = 51.164 \pm 0.007$\,km\,s$^{-1}$ and $K_2 =  49.106\pm 0.010$\,km\,s$^{-1}$.

 The parameters of interest for the calculation of the masses and radii are $r_1= R_1/a$, $r_2= R_2/a$, $\sin i$ and $e$. From Table~\ref{geompar} and \ref{radpar} we see that the sample standard deviation across the analysis runs is similar to the typical error estimate for each parameter. We therefore decided to use the mean value of each parameter given in Table~\ref{geompar} to calculate the masses and radii of the stars. From Fig.~\ref{ResultsLCFit} we see that some of these parameters are correlated so we used a Monte-Carlo approach to calculate errors and covariances for the masses and radii. We used {\sc emcee} to obtain 50,000 randomly sampled values from a multi-variate normal distribution with the same mean and covariance matrix as the input values of $r_1$, $r_2$, $\sin i$ and $e$. These 50,000 sets of parameters were paired with the same number of $K_1$ and $K_2$ values from two independent normal distributions. For the orbital period we used the value $P=24.5924$\,d and assumed that the error in this value is negligible.  We then used equations and constants from \citet{2016AJ....152...41P} to calculate the masses and radii of the stars in nominal solar units for every set of randomly sampled parameters. We also calculated the mean stellar density and surface gravity for each star. The mean and sample standard deviation of these random samples are given in Table~\ref{MassRadiusTable} and the parameter correlation plots for the PPDs are shown in Fig.~\ref{MassRadiusPlot}.

   \begin{figure}
   \centering
   \includegraphics[width=0.47\textwidth]{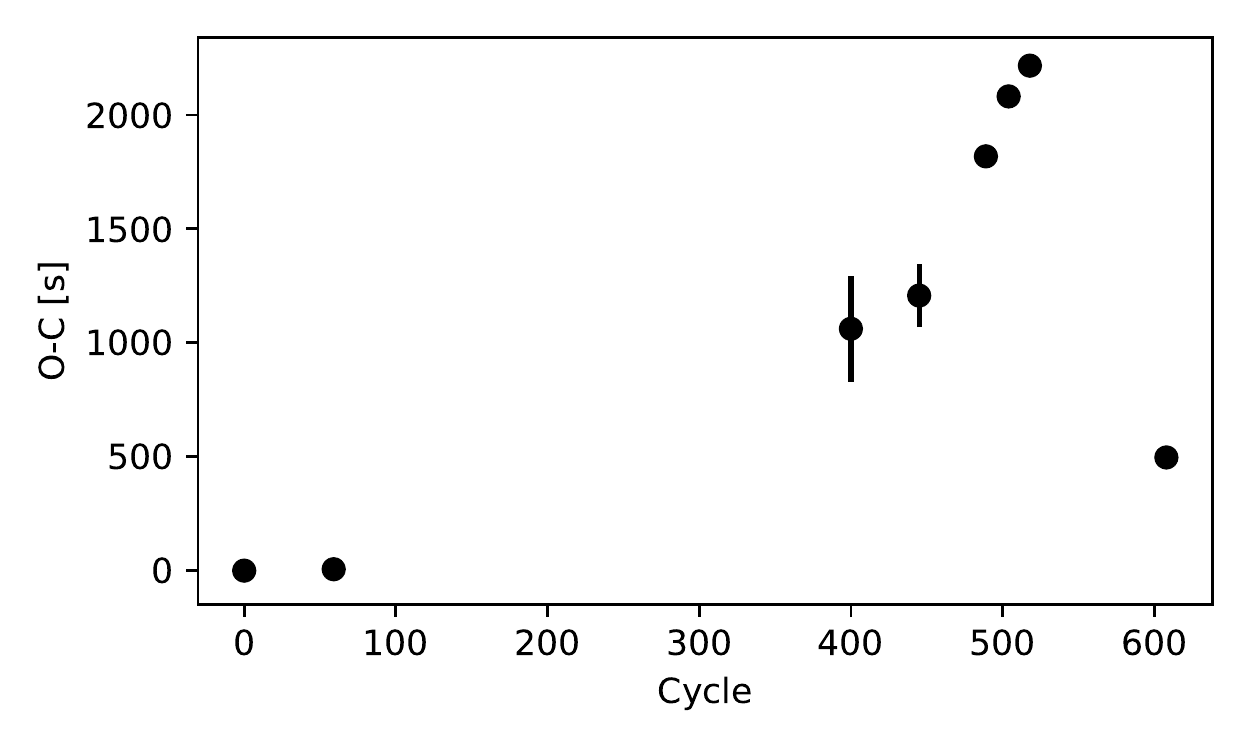}
      \caption{Residuals from the ephemeris of \citet{1984ApJ...282..748H} for times of mid-primary eclipse converted to BJD$_{\rm TDB}$. The ephemeris on this time system is BJD$_{\rm TDB} = 2443410.6891 + 24.592325\cdot E$, where $E$ is the cycle number.}
         \label{EphemPlot}
   \end{figure}
%

\begin{table}
\caption[]{Additional parameters derived from the analysis of the TESS light curve of AI~Phe and the root-mean-square (rms) of the residuals from the best fits. A $\star$ symbol in the final column denotes runs that only included data in the region of the eclipses in the fit. }
\label{radpar}
\begin{center}
  \begin{tabular}{@{}llllcc}
\hline
 \noalign{\smallskip}
\multicolumn{1}{@{}l}{Run} &
 \multicolumn{1}{l}{$e$} &
 \multicolumn{1}{l}{$\omega\:[^{\circ}]$} &
 \multicolumn{1}{l}{$\ell_3$} &
 \multicolumn{1}{l}{rms} \\
  & & & & \multicolumn{1}{l}{[ppm]} 
 \\
\noalign{\smallskip}
\hline
\noalign{\smallskip}
A  & 0.1884(7)   & 110.23(8)   & 0.006(2)   & 356 & $\star$ \\
B  & 0.1871(11)  & 110.39(13)  & 0.010(2)   & 416 & \\
C  & 0.1878(6)   & 110.30(6)   & 0.008(2)   & 368 & \\
D  & 0.1883(10)  & 110.24(11)  & 0.007(2)   & 368 & \\
E  & 0.1870(7)   & 110.40(1)   & 0.007(1)   & 428 & $\star$ \\
F  &  0.1866(1)           &  110.45(1)    & 0.006(4)   & 390 & $\star$ \\
G  & 0.1859(4)   & 110.54(5)   & 0.007(2)   & 372 & \\
H  & 0.1866(12)  & 110.40(16)  & 0.025(25)  & 353 & $\star$ \\
I  & 0.1875(15)  & 110.41(12)  & 0.025(25)  & 353 & $\star$ \\
J  & 0.1862(11)  & 110.48(12)  & 0.025(25)  & 353 & $\star$ \\
K  & 0.1881(8)   & 110.27(9)   & 0.0053(25) & 381 & $\star$ \\
L  & 0.1894(17)  & 110.12(19)  & 0.0072(35) & 381 & $\star$ \\
S  & 0.1881(7)   & 110.27(8)   & 0.005(2)   & 364 & $\star$ \\
\noalign{\smallskip}
\hline
\noalign{\smallskip}
-- & 0.1875(9)   & 110.34(11)   & 0.011(8)   & -- & -- \\

\noalign{\smallskip}
\hline
\end{tabular}
\end{center}
\end{table}
%

\begin{table}
\caption[]{Masses, radii and derived parameters for the stars in AI~Phe.  Surface gravity, $g$, is given in cgs units, $\rho$ is the mean stellar density and $C(x,y)$ is the covariance of $x$ and $y$. }
\label{MassRadiusTable}
\begin{center}
  \begin{tabular}{@{}lrr}
\hline
\noalign{\smallskip}
 \multicolumn{1}{@{}l}{Parameter} &
 \multicolumn{1}{l}{Value} &
 \multicolumn{1}{r}{} \\
\noalign{\smallskip}
\hline
\noalign{\smallskip}
$M_1/{\mathcal M^{\rm N}_{\odot}}$&1.1938 $\pm$ 0.0008 &[0.07\%] \\
\noalign{\smallskip}
$M_2/{\mathcal M^{\rm N}_{\odot}}$&1.2438 $\pm$ 0.0008 &[0.06\%] \\
\noalign{\smallskip}
$R_1/{\mathcal R^{\rm N}_{\odot}}$&1.8050 $\pm$ 0.0022 &[0.12\%] \\
\noalign{\smallskip}
$R_2/{\mathcal R^{\rm N}_{\odot}}$&2.9332 $\pm$ 0.0023 &[0.08\%] \\
\noalign{\smallskip}
$\rho_1/{\rho^{\rm N}_{\odot}}$  &0.20299 $\pm$ 0.00076 &[0.38\%] \\
\noalign{\smallskip}
$\rho_2/{\rho^{\rm N}_{\odot}}$  & 0.04928 $\pm$ 0.00014 &[0.29\%] \\
\noalign{\smallskip}
$\log g_1$   & 4.0020 $\pm$ 0.0011 & [0.03\%]  \\
\noalign{\smallskip}
$\log g_2$   & 3.5981 $\pm$ 0.0009 & [0.03\%] \\
\noalign{\smallskip}
\hline\noalign{\smallskip}
$C(M_1/{\mathcal M^{\rm N}_{\odot}},\:M_2/{\mathcal M^{\rm N}_{\odot}})$ & $5.89\times10^{-7}$ \\
\noalign{\smallskip}
$C(M_1/{\mathcal M^{\rm N}_{\odot}},\:R_1/{\mathcal R^{\rm N}_{\odot}})$ & $-1.51\times10^{-7}$ \\
\noalign{\smallskip}
$C(M_2/{\mathcal M^{\rm N}_{\odot}},\:R_2/{\mathcal R^{\rm N}_{\odot}})$ & $-1.04\times10^{-6}$ \\
\noalign{\smallskip}
$C(R_1/{\mathcal R^{\rm N}_{\odot}},\:R_2/{\mathcal R^{\rm N}_{\odot}})$ & $2.06\times10^{-6}$ \\
\noalign{\smallskip}
\hline
\end{tabular}
\end{center}
\end{table}

\section{Discussion}
\subsection{Consistency checks}
A valuable test for the presence of unrecognised systematic errors is to compare values determined independently from separate data sets. In particular, $e$ and $\omega$ are determined from the spectroscopic orbit and light curve independently to good precision. The values of $e\cos \omega$ and  $e\sin \omega$ from the spectroscopic orbits of \citet{2018MNRAS.478.1942S} and  \citet{2019A&A...632A..31G} are compared to the values obtained from the light curve in Fig.~\ref{ResultsLCFit}. A detailed statistical comparison of these values is not straightforward because there will be some covariance between these parameters from the spectroscopic orbit, and the values obtained from the light curve are affected by systematic errors. Nevertheless, by-eye there appears to be good agreement between these values obtained from spectroscopy and photometry. The values of   $e\cos \omega$ and  $e\sin \omega$ from \citet{2009MNRAS.400..969H} are also consistent with the values from the light curve analysis and the two other spectroscopic orbits, although with significantly larger error bars.

The inclination of the binary orbit determined from the astrometric orbit by \citet{2019A&A...632A..31G} is also consistent with the value of $i$ determined from the light curve, once the ambiguity in the sign of $\cos i$ determined from the light curve is accounted for. This is a less stringent check because of the relatively large error in the astrometric value.

One check that is not yet possible is to compare these results to those from the analysis light curves of similar quality observed at different times. Additional observations of AI~Phe that will enable this test to be done are expected during Cycle 3 of the TESS mission. A less stringent test but one that is nevertheless important is to check that the detrending of the light curve has not biased the results. This can be done by comparing the values obtained for Run A and Run S in which we analysed the {\tt PDCSAP\_FLUX} data and the {\tt SAP\_FLUX} data in the same way. The results from these two runs are consistent within their estimated errorbars so we conclude that the de-trending has not biased the results in this case.

The parameters from this study are within about 2 standard deviations of the quoted value and errors from previous studies. This level of agreement is quite satisfactory given the difficulties in calculating the orbital phase for data from different observing seasons caused by the long and variable orbital period of AI~Phe. 
   
\subsection{Limb-darkening}
  \citet{2018A&A...616A..39M} showed that the parameters $h_1 =
I_{\lambda}(\nicefrac{1}{2})$ and $h_2 = I_{\lambda}(\nicefrac{1}{2}) - I_{\lambda}(0)$ are useful for comparing the limb-darkening measured from light curves of transiting exoplanets to model predictions because they are not strongly correlated with one another. Here $I_{\lambda}(\mu)$ is the surface brightness relative to the value at the centre of the stellar disc, and $\mu$ is the cosine of the angle between the line of sight and the normal vector to the stellar surface. The power-2 limb-darkening law used by \citet{2018A&A...616A..39M} is defined by $I_{\lambda}(\mu) = 1-c\left(1-\mu^{\alpha}\right)$, so $h_1 = 1-c\left(1-2^{-\alpha}\right)$ and $h_2 = c2^{-\alpha}$. From Run A we obtain the values $h_{\rm 1,F}=0.8167 \pm 0.0036$ and  $h_{\rm 2,F}=0.61 \pm 0.11$ for star 1 (the F7\,V star). Note that these parameters are constrained by Gaussian priors with standard deviations of 0.05 and 0.25 for $h_1$ and $h_2$, respectively. If we assume $T_{\rm eff,F} = 6310\pm100$\,K then interpolation in  Table 2 of \citet{2018A&A...616A..39M} gives $h_{\rm 1,F}= 0.818 \pm 0.004$ and $h_{\rm 2,F} = 0.418\pm 0.003$. The agreement between the observed and model values of $h_{\rm 1,F}$ is remarkable and is consistent with the conclusion from \citet{2018A&A...616A..39M} that the values of $h_1$ derived from the \textsc{Stagger}-grid stellar atmosphere models \citep{2015A&A...573A..90M} are accurate to about $\pm0.01$ for dwarf stars with $T_{\rm eff} \approx 6000$\,K. The value of $h_{\rm 2,F}$ does not fit with the trend observed in \citet{2018A&A...616A..39M} for the observed values $h_2$ to be lower than the predicted values by about 0.05. However, the effect of $h_{\rm 2,F}$  on the light curve of AI~Phe is extremely subtle so a much more careful analysis of the systematic error in this value would be needed before drawing any firm conclusions. For the K0\,IV star we obtain the values
$h_{\rm 1,K} = 0.761 \pm 0.049$ and $h_{\rm 2,K} = 0.63 \pm 0.12$ from Run A. The error on $h_{\rm 1,K}$ is almost equal to the standard deviation of the Gaussian prior on this value, i.e., there is no constraint on this parameter from the light curve. 

 We also compared the quadratic limb darkening coefficients, $u_1$ and $u_2$, obtained from Run D to the tabulated values calculated from stellar model atmospheres by \citet{2011A&A...529A..75C}. The values from the light curve analysis are  $u_1 = 0.252 \pm 0.024$, $u_2 = 0.246 \pm 0.053$ for the F7\,V star and  $u_1 = 0.409 \pm 0.036$, $u_2 = 0.237 \pm 0.075$ for the K0\,IV star. Interpolating using parameters  $T_{\rm eff,1} = 6310$\,K, T$_{\rm eff,2} = 5010$\,K, $\log g_1 = 4.00$, $\log g_2 = 3.60$, and [Fe/H]$ = -0.10$, we obtain  $u_1= 0.210$, $u_2 = 0.307$ for the F7\,V star and 
  $u_1= 0.392$, $u_2 = 0.219$ for the K0\,IV star, which are consistent with the observed values to within 0.1.
 
 The coefficients of the cubic limb-darkening law were found to be much less well constrained. Indeed, for a few of the simulated light curves used in the residual-permutation analysis the best-fit light curve required coefficients that are not physically realistic. This may explain why the error estimates from Run L are noticeably larger than those from other analysis methods. 
 
\subsection{Times of mid-eclipse and third light.}
   We define the time of mid-eclipse to be the time when the angular separation on the sky between the centres of the two stellar discs is at a minimum \citep{1992AJ....104.2213L}. The times of mid-eclipse from Table~\ref{geompar} tend to fall into two groups -- those near  2458362.82850 and another group near 2458362.82834. The times of mid-eclipse near the latter value have, with the exception of Run G, been derived using either light curve codes based on the {\sc ebop} algorithm or the Wilson-Devinney algorithm. Inspection of the source code reveals that both these algorithms use the approximation $i=90^{\circ}$ to calculate the mean anomaly at mid-primary eclipse.  A similar approach was also used for Run G.  For AI~Phe this approximation results in an offset of 0.00013 days between the predicted and observed time of mid-eclipse. If we apply this correction then the unweighted mean time of mid-primary eclipse is 2458362.82850 with standard deviation of 0.00002 days.   Fig.~\ref{EphemPlot} shows this time of minimum as a residual from the ephemeris  of \citet{1984ApJ...282..748H}. Note that all published times of minimum were converted to Barycentric Dynamical Time (TDB) for consistency with the timestamps used in the TESS archive data products \citep{tessdata} prior to calculating the residuals shown in this figure. There has been a quite dramatic shift in the time of primary eclipse from this linear ephemeris since the observations obtained with the WASP instrument reported in \citet{2016A&A...591A.124K}. This is likely to be due to the periastron passage of the putative third body in the system some time between JD 2456000 and 2458000.
 
 A complete dynamical analysis of the AI~Phe system is beyond the scope of this paper, but may be worthwhile in order to better understand whether the third body might be a stellar remnant (e.g., white dwarf) or a low-mass star. A useful constraint in such an analysis is the observation that this third body is very faint compared to the eclipsing components. We used the broadening function method \citep{1992AJ....104.1968R} to look for any sign of third light in the HARPS spectra of AI Phe analysed by \citet{2019A&A...632A..31G}. If we assume that the third light is due to an M-dwarf with T$_{\rm eff}\sim 3000$\,K then this star contributes no more than about 1\% of the optical flux. The interferometric observations reported in Gallenne et al. also enable us to put a limit of ${\rm H}>10.5$ for the magnitude of any resolved companions within 100\,mas of the eclipsing binary. A third-light contribution of $\ell_3=0.7\%$ in the TESS band would correspond to a K9V star at a distance of AI~Phe. Such a star would have a H-band magnitude $H\approx 11.3$, which is consistent with this non-detection. Comparing the results for $\ell_3$ in Table~\ref{radpar} to our estimate of 0.5\% for the contamination from the faint star seen Fig.~\ref{TESS_Aperture_Overlay_Skymapper} we see that the third body probably contributes no more than about 0.3\% of the flux in the TESS band. However, it should be noted that it is not clear how robust the estimate of $\ell_3$ from the light curve analysis is. For Runs H, I and J we find there is essentially no constraint on $\ell_3$ within the uniform prior over the range $[0,0.05]$ imposed during the MCMC analysis. This is in contrast to other runs which typically find $\ell_3\approx 0.007\pm 0.002$. The reason for this difference is not clear but appears not to have a significant impact on the results so we have not investigated  this issue further.

\subsection{Error estimates}
  The main parameters from the light curve analysis of interest for the calculation of the masses and radii of the stars are $r_1$, $r_2$ and $i$. The standard error estimates for these parameters in Table~\ref{geompar} vary by a factor of 5 or more. Given the level of agreement between the different methods it is clear that the error estimates produced by some methods are underestimated, sometimes severely so. This is a well-known problem for light curve codes that calculate so-called ``formal'' error estimates based on a quadratic approximation for the dependence of $\chi^2$ on the model parameters in the region of the best fit, e.g. the Wilson-Devinney model used for Run E.   Indeed, the Cramer-Rao theorem states that any unbiased estimator for the parameters will deliver a covariance matrix on the parameters that is no better than this \citep{2009arXiv0901.0721A}.

\section{Conclusions}

 The robustness of our results for AI~Phe is partly due to the fact that the primary eclipse is total. This gives a direct measurement of the flux ratio for the binary  from the depth of the eclipse where one star is completed occulted \citep{1912ApJ....35..315R}. This key parameter can be strongly correlated with the radius ratio and inclination for systems with partial eclipses, particularly if the eclipses are shallow. We expect that the analysis of TESS light curves for other eclipsing binaries with deep or total eclipses will, in general, be similarly robust. For systems with partial eclipses we recommend that the dependence of the results on assumptions such as the limb-darkening model, third light contribution and detrending method used in the analysis should be careful examined and presented along with the parameters of interest. 
 
 The comparison of results obtained using different methods and analysed independently by different researchers has been illuminating. The results for the light curve analysis presented here are typically not the first results that were produced by each analyst. An initial comparison of our results revealed a number of issues, some of which are described above. These issues generally have a small effect on the results but even small biases can be significant when working  with space-based photometry. We strongly recommend that independent analysis using two or more methods are carried out and reported when using space-based photometry to characterise eclipsing binaries to high precision ($\la 0.5$\%). 
 
 The value of inspecting the residuals from the best light curve fit should not be underestimated. We recommend that these residuals should always be clearly shown in any published analysis of an eclipsing binary as has been done here in Fig.~\ref{MaxtedLCFit}. The residuals from the other best-fit light curve models presented here (with the exception of Run E) are equally good, i.e., there is no noticeable increase in the scatter of the residuals through the eclipse cf. the out-of-eclipse phases, and no trends within the residuals indicative of a poor fit to the depth or shape of the eclipse. If this is not the case then some level of systematic error in the parameters derived from the model must be expected. This was found to be the case for AI~Phe if we used a linear limb-darkening law to model the light curve. 
 
  The preferred method for estimating errors on the light curve parameters is to analyse multiple independent data sets. If multiple orbits of the binary systems have been observed by one instrument this is easily achieved by analysing subsets of the data \citep[e.g.][]{2018RNAAS...2...39M, 2019MNRAS.484..451H}. Taking the standard error of the mean across subsets is a straightforward and robust way to estimate the error on each parameter if at least 3 (preferably 4) subsets can be created. This may also mitigate systematic errors in the results from imperfect model fits in some cases if the poor fit is caused by an effect that ``averages out'' over multiple eclipses. For long-period binaries such as AI~Phe where the light curve only covers one or two orbital cycles, Monte-Carlo methods such as {\tt emcee} or residual permutation can give realistic error estimates if used correctly,  as can be seen in Fig.~\ref{ResultsLCFit}. The  ``formal'' errors obtained from the covariance matrix are certainly underestimates of the true uncertainty and should not be used.

The agreement between the masses obtained from spectra of AI~Phe obtained at high resolution (R$>$60,000) with good S/N and good phase coverage is very good. This is an ideal case -- both stars rotate slowly and so have narrow spectral lines and low levels of stellar activity. The accuracy of the masses may not be so good for rapidly rotating or magnetically-active stars, particularly if there is a lack of spectra covering the key orbital phases when the stars are near their maximum and minimum radial velocities.

Differences between the various limb-darkening laws we have used is a significant contribution to the  error in the radius in our analysis. It is not clear from this analysis which is the preferred limb-darkening law to use, although the linear limb darkening approximation is certainly not accurate enough. We can expect progress to be made in our understanding of limb darkening for cool stars over coming years because the same issue also affects observational studies of transiting exoplanets. This problem can be tackled using high-quality light curves for transiting exoplanet systems and suitable eclipsing binaries from TESS and other space-based instrumentation.

\section*{Acknowledgements}
This paper includes data collected by the TESS mission. Funding for the TESS
mission is provided by the NASA Explorer Program. PM acknowledges support from
Science and Technology Facilities Council, research grant number ST/M001040/1.
KGH acknowledges support provided by the Polish National Science Center
through grant no. 2016/21/B/ST9/01613. AP acknowledges NASA support via grant
\#80NSSC18K0417 and NSF support via grant \#1909109. This work has received
funding from the European Research Council (ERC) under the European Union's
Horizon 2020 research and innovation programme (CartographY GA. 804752).  GRD,
WB, and WJC acknowledge the support of the UK Science and Technology
Facilities Council (STFC). The national facility capability for SkyMapper has
been funded through ARC LIEF grant LE130100104 from the Australian Research
Council, awarded to the University of Sydney, the Australian National
University, Swinburne University of Technology, the University of Queensland,
the University of Western Australia, the University of Melbourne, Curtin
University of Technology, Monash University and the Australian Astronomical
Observatory. SkyMapper is owned and operated by The Australian National
University's Research School of Astronomy and Astrophysics. The survey data
were processed and provided by the SkyMapper Team at ANU. The SkyMapper node
of the All-Sky Virtual Observatory (ASVO) is hosted at the National
Computational Infrastructure (NCI). Development and support the SkyMapper node
of the ASVO has been funded in part by Astronomy Australia Limited (AAL) and
the Australian Government through the Commonwealth's Education Investment Fund
(EIF) and National Collaborative Research Infrastructure Strategy (NCRIS),
particularly the National eResearch Collaboration Tools and Resources (NeCTAR)
and the Australian National Data Service Projects (ANDS). CJ recognises
support form the European Research Council (ERC) under the European Union's
Horizon 2020 research and innovation programme (grant agreement
N$^\circ$670519: MAMSIE), and from the Research Foundation Flanders (FWO)
under grant G0A2917N (BlackGEM). Parameter correlation plots were generated
using the {\tt python} package {\tt corner}  \citep{corner}. We would like to
thank the anonymous referee for their constructive and timely comments on the
manuscript.


\bibliographystyle{mnras} 
\bibliography{aiphe}


\bsp  
\label{lastpage}
\end{document}